\documentstyle[epsfig]{aipproc}

\begin{document}
\title{Determination of the Gaugino Mass Parameter $M_1$ in Different Linear Collider Modes\thanks{Presented by Claus Bl\"ochinger.}}

\author{C. Bl\"ochinger$^*$,  H. Fraas$^*$, T. Mayer$^*$, G. Moortgat-Pick$^{\dagger}$}
\address{$^*$Institut f\"ur Theoretische Physik und Astrophysik, Universit\"at W\"urzburg, Am Hubland, 97074 W\"urzburg, Germany\thanks{Work supported by the Deutsche Forschungsgemeinschaft, contract FR 1064/4-1 and the Bundesministerium f\"ur Bildung und Forschung, contract 05 HT9WWa 9.}\\
$^{\dagger}$ DESY, Deutsches Elektronen-Synchrotron, D-22603 Hamburg, Germany}

\maketitle

\begin{abstract}
We study the different linear collider modes  with regard to the determination
of the gaugino mass parameter $M_1$. In a specific mSUGRA inspired scenario
we compare four processes with polarized beams:
(a) $e^+e^- \rightarrow \tilde{\chi}_1^0\tilde{\chi}_2^0 \rightarrow \tilde{\chi}_1^0\tilde{\chi}_1^0e^+e^-$, (b) $e^-\gamma \rightarrow \tilde{\chi}_1^0\tilde{e}_{L/R}$\\
$\rightarrow \tilde{\chi}_1^0\tilde{\chi}_1^0e^-$, (c) $\gamma\gamma \rightarrow \tilde{\chi}_1^+\tilde{\chi}_1^- \rightarrow \tilde{\chi}_1^0\tilde{\chi}_1^0e^+e^-\nu_e\bar{\nu}_e$, (d) $e^-e^- \rightarrow \tilde{e}_{L/R}^-\tilde{e}_{L/R}^- \rightarrow \tilde{\chi}_1^0\tilde{\chi}_1^0e^-e^-$.
\end{abstract}

\section*{Introduction}

After discovering supersymmetry it will be particularly important to 
measure the parameters of the underlying supersymmetric model at a linear
collider \cite{LC1}. In the Minimal Supersymmetric Standard Model (MSSM) the
neutralino sector depends on the $U(1)$ and $SU(2)$ gaugino
mass parameters $M_1$ and $M_2$, the higgsino mass parameter $\mu$ and the 
ratio $\tan\beta$ of the vacuum expectation values of the Higgs fields. 
For the gaugino mass parameters usually the GUT relation 
$M_1=5/3\tan^2\theta_W\times M_2 $ is assumed. We relax this relation and 
present for processes in the four different 
linear collider modes a comparative  study of the $M_1$ dependence of the
production of supersymmetric particles with subsequent leptonic decay. Beam
polarization and complete spin correlations between production and decay are
included. We discuss the possibility to determine 
the gaugino mass parameter $M_1$ and, combined with measurements 
from the chargino sector, to test  the GUT relation between $M_1$ and
$M_2$. In the following a mSUGRA inspired scenario is assumed with 
$M_2=152$ GeV, $\mu=316$ GeV and $\tan\beta=3$, which leads to a gaugino-like
LSP $\tilde{\chi}_1^0$ \cite{scenario}.

\section*{Neutralino Pair Production in $e^+e^-$ Annihilation}

In fig. 1a the total cross sections $\sigma_e$ and in fig. 1b
the forward-backward asymmetries $A_{FB}$ of the decay electron of the process 
$e^+e^- \rightarrow \tilde{\chi}_1^0\tilde{\chi}_2^0 \rightarrow \tilde{\chi}_1^0\tilde{\chi}_1^0e^+e^-$ 
for $\sqrt{s}=m_{\tilde{\chi}_1^0}+m_{\tilde{\chi}_2^0}+30$ GeV and selectron 
masses $m_{\tilde{e}_R}=161$ GeV and $m_{\tilde{e}_L}=176$ GeV are shown 
for various configurations of beam polarization \cite{Gudi1}. 
Beam polarization strongly enlarges the sensitivity of both observables to 
$M_1$. The cross sections allow to constrain $M_1$ with ambiguities, 
which can be resolved by measuring $A_{FB}$.
Since $\vert P_{e^-}\vert = 85 \%$, $\vert P_{e^+}\vert = 60 \%$ enhances
$\sigma_e$ by about a factor 3, the statistical errors for the observables 
are reduced by more than 40\% \cite{Gudi2}.
As can be seen in \cite{Gudi1} $M_1$ can also be determined by
measuring polarization asymmetries of the cross section.

\begin{figure}[h]
\label{e+e-}
\centering
\begin{picture}(15,5.5)
\put(-1.5,-1.1){\includegraphics{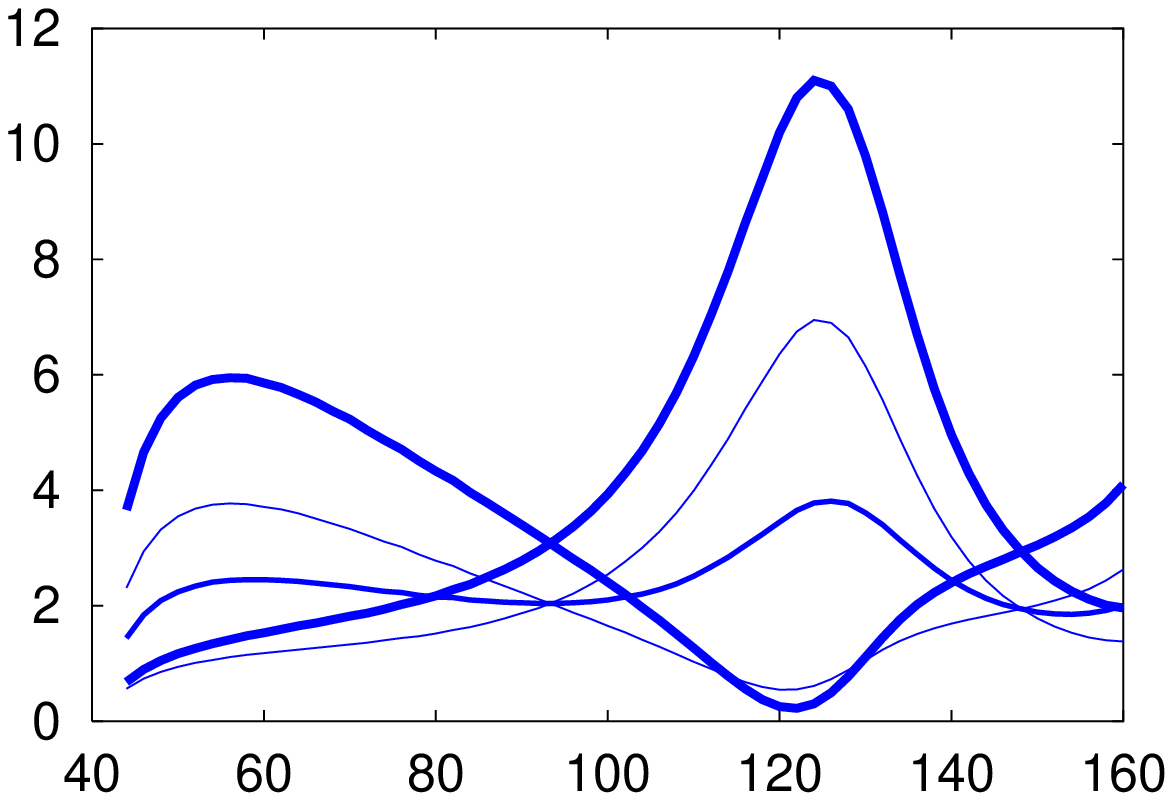}}
\put(6.33,0.05){{\tiny $M_1$/GeV}}
\put(0.2,5.35){{\tiny $\sigma_{e}$/fb}}
\put(1.3,3.0){{\tiny $\left(-+\right)$}}
\put(1.3,2.2){{\tiny $\left(-0\right)$}}
\put(4.9,3.3){{\tiny $\left(+0\right)$}}
\put(4.9,2.2){{\tiny $\left(00\right)$}}
\put(4.9,4.75){{\tiny $\left(+-\right)$}}
\put(6.0,-1.1){\includegraphics{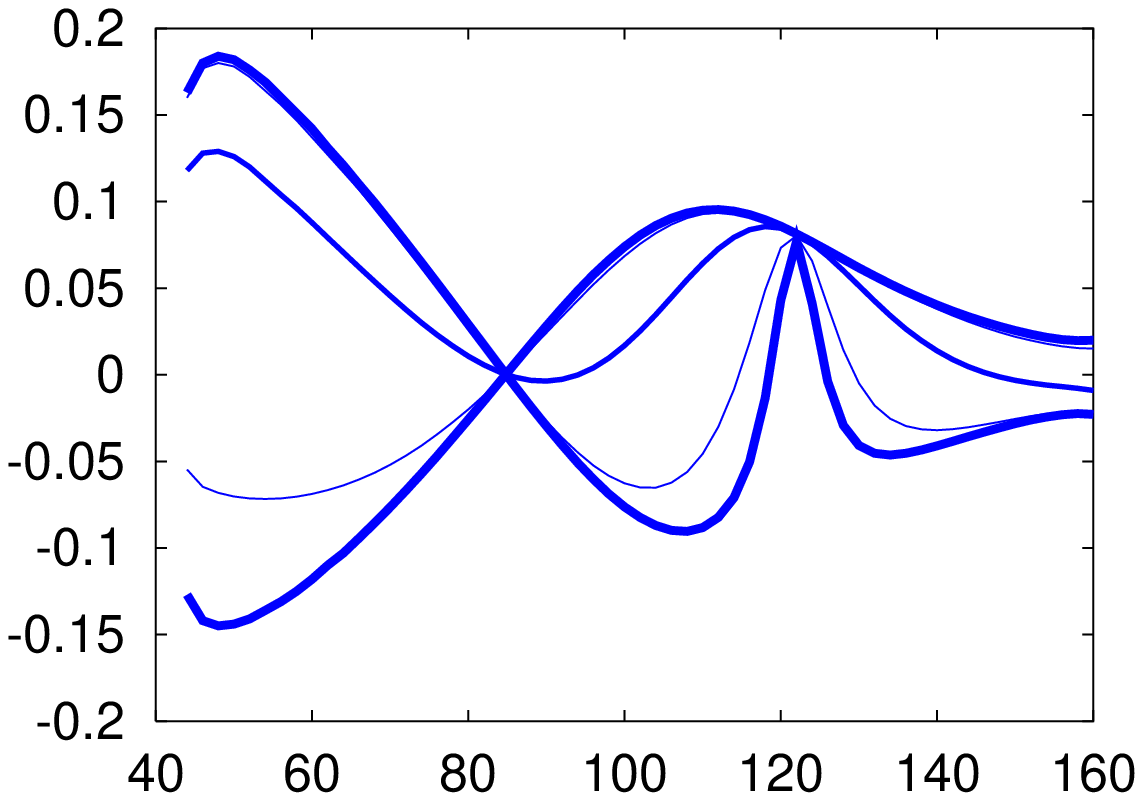}}
\put(13.8,0.05){{\tiny $M_1$/GeV}}
\put(8.4,5.35){{\tiny $A_{FB}$}}
\put(9.5,4.7){{\tiny $\left(-+\right)\approx \left(-0\right)$}}
\put(11.9,1.65){{\tiny $\left(-+\right)$}}
\put(9.0,2.35){{\tiny $\left(+0\right)$}}
\put(9.0,3.8){{\tiny $\left(00\right)$}}
\put(9.0,1.05){{\tiny $\left(+-\right)$}}
\put(11.5,2.4){{\tiny $\left(-0\right)$}}
\put(12.5,3.9){{\tiny $\left(+-\right)\approx \left(+0\right)$}}
\put(3.7,5.35){{\tiny (a)}}
\put(11.45,5.35){{\tiny (b)}}
\end{picture}
\caption{Total cross sections $\sigma_{e}$ (a) and forward-backward asymmetries $A_{FB}$ (b) of the decay electron for $e^+e^- \rightarrow \tilde{\chi}_1^0\tilde{\chi}_2^0 \rightarrow \tilde{\chi}_1^0\tilde{\chi}_1^0e^+e^-$ for 
unpolarized beams and for longitudinal electron (positron) polarizations $P_{e^-}=\pm 85\%$ ($P_{e^+}=\pm 60\%$)  at $\sqrt{s}=m_{\tilde{\chi}_1^0}+m_{\tilde{\chi}_2^0}+30$ GeV. The polarization is denoted as $(sign(P_{e^-}),sign(P_{e^+}))$.}
\end{figure}

\section*{Associated Production of Selectrons and Neutralinos in $e^-\gamma$ Scattering}

Due to the large cross sections for a gaugino-like $\tilde{\chi}_1^0$, the 
process $e^-\gamma \rightarrow \tilde{\chi}_1^0\tilde{e}_{L/R} \rightarrow \tilde{\chi}_1^0\tilde{\chi}_1^0e^-$ offers an interesting possibility to access 
the gaugino mass parameter $M_1$. High energetic photon beams are
obtained by Compton backscattering of laser beams off one of the electron
beams of a linear collider. Both the energy spectrum and the mean helicity
of the high energetic photons sensitively depend on the longitudinal 
polarization $P_{e2}$ of the converted electrons and the circular polarization
$\lambda_L$ of the laser photons \cite{PC1}. In addition the longitudinal
polarization $P_{e1}$ of the colliding electron beam can be used to enhance 
the contribution of right and left selectrons, respectively. 
The total cross section $\sigma_{ee}$ in the laboratory frame, obtained by
convoluting the $e\gamma$ cross section with the energy distribution of the high energetic photons, is shown in fig. 2 for $\sqrt{s_{ee}}=500$ GeV and 
different polarizations of the electron beams and the laser photons. For all
polarization configurations the cross sections in our scenario are larger
than those for $e^+e^- \rightarrow \tilde{\chi}_1^0\tilde{\chi}_2^0 \rightarrow \tilde{\chi}_1^0\tilde{\chi}_1^0e^+e^-$, fig. 1. For 40 GeV $<M_1<$ 170 GeV
high cross sections are obtained for $P_{e1}=P_{e2}=80\%$ and 
for both right and left circularly polarized laser photons, 
$\lambda_L=\pm 100\%$ (fig. 2a), so that in this region $M_1$ can be 
measured with 
small statistical errors. Although being somewhat smaller the 
cross sections for $P_{e1}=P_{e2}=-80\%$ and $\lambda_L=\pm 100\%$,
fig. 2b, significantly depend on $M_1$ up to 300 GeV. Ambiguities in the 
cross sections can easily be resolved by measuring polarization 
asymmetries \cite{bloechi1}.

\begin{figure}[hbt]
\label{e-gamma}
\centering
\begin{picture}(15,5.2)
\put(-1.1,-3.6){\includegraphics{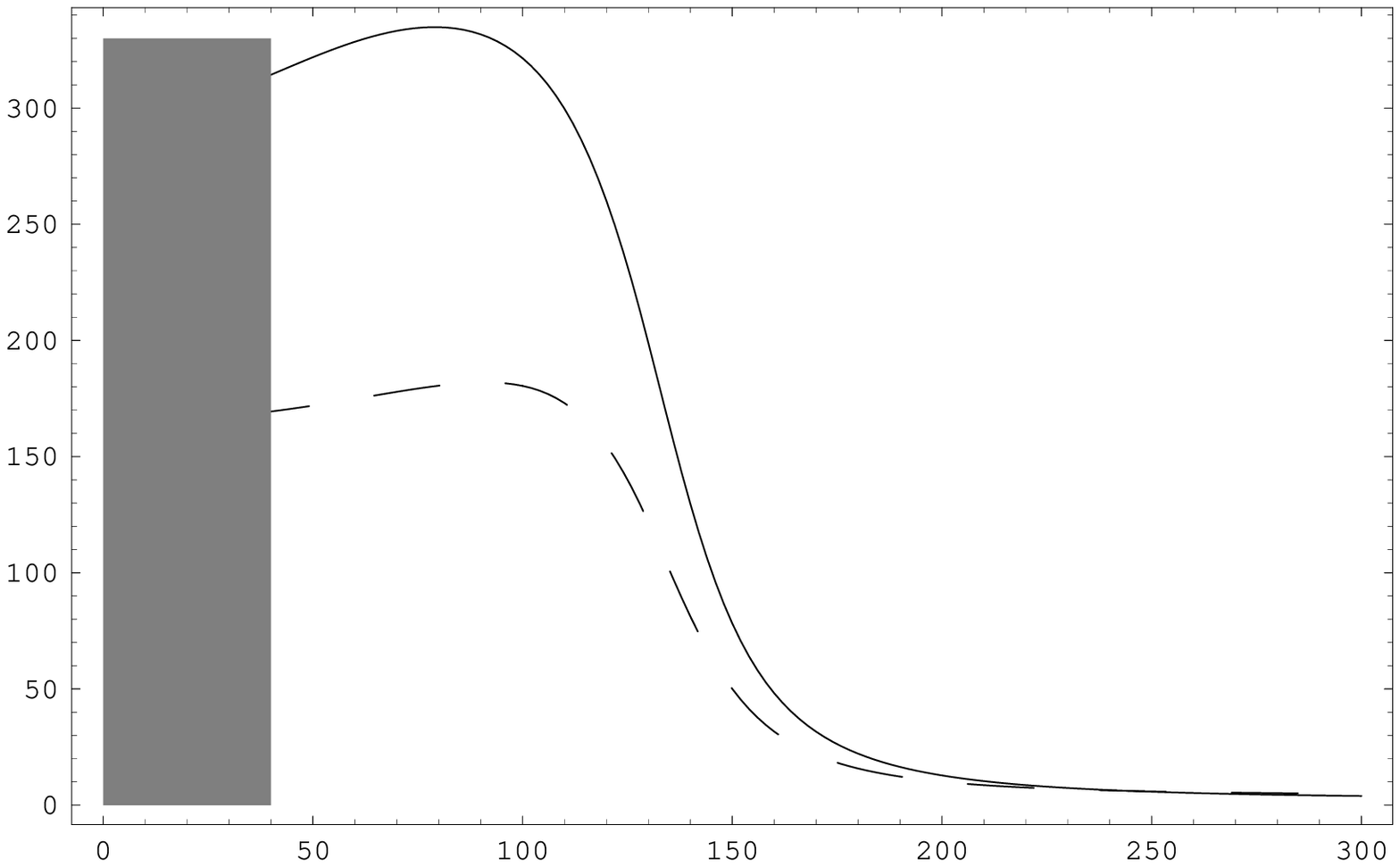}}
\put(6.33,0.05){{\tiny $M_1$/GeV}}
\put(0.05,5.){{\tiny $\sigma_{ee}$/fb}}
\put(6.6,-3.6){\includegraphics{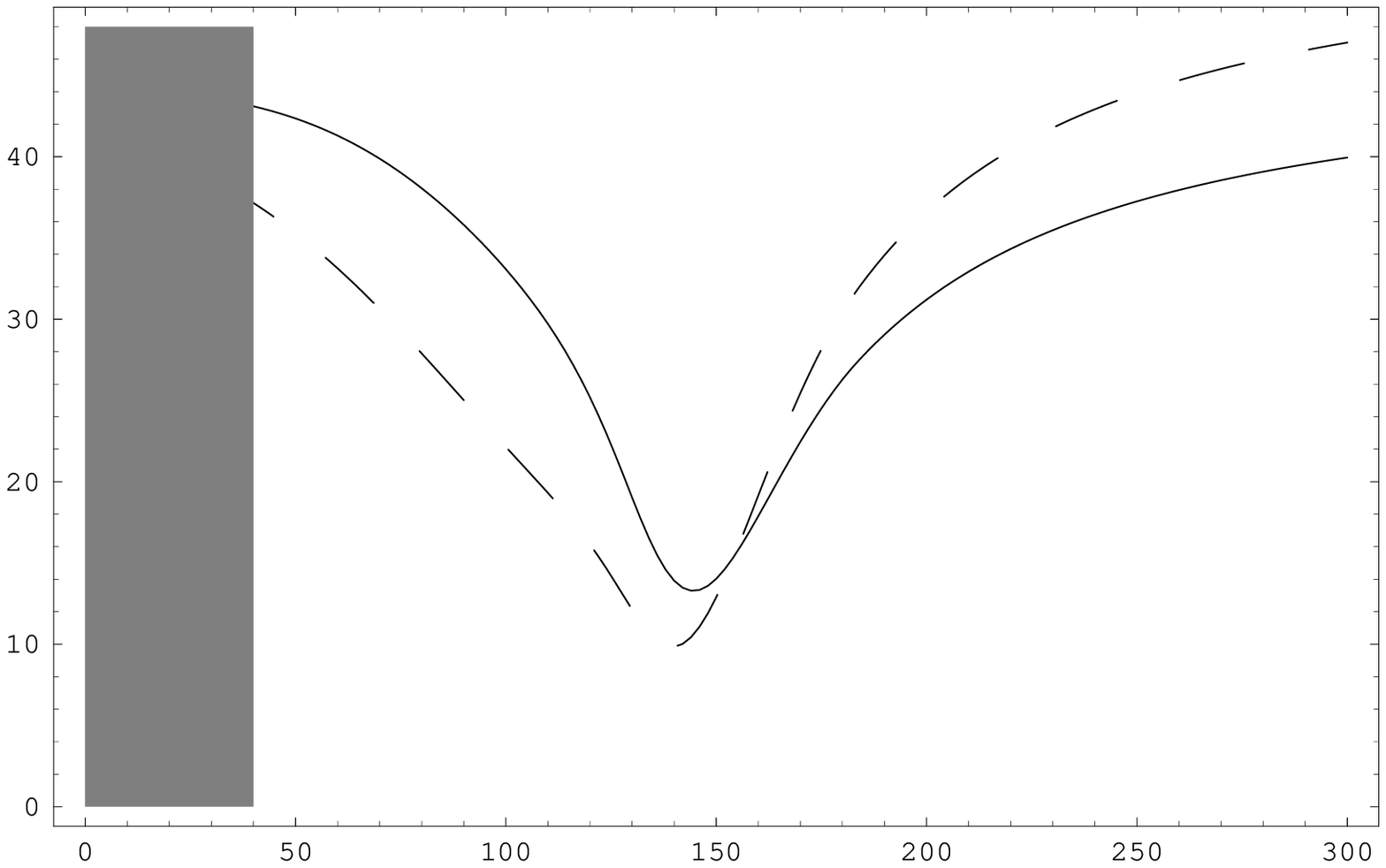}}
\put(13.8,0.05){{\tiny $M_1$/GeV}}
\put(7.6,5.){{\tiny $\sigma_{ee}$/fb}}
\put(3.,4.5){{\tiny $\left(+++\right)$}}
\put(1.6,3.1){{\tiny $\left(++-\right)$}}
\put(13.5,3.5){{\tiny $\left(--+\right)$}}
\put(13.,4.6){{\tiny $\left(---\right)$}}
\put(3.7,5.0){{\tiny (a)}}
\put(11.35,5.0){{\tiny (b)}}
\end{picture}
\caption{Total cross sections $\sigma_{ee}$ of $e^-\gamma \rightarrow \tilde{\chi}_1^0\tilde{e}_{L/R} \rightarrow \tilde{\chi}_1^0\tilde{\chi}_1^0e^-$ for $\sqrt{s_{ee}}=500$ GeV, $m_{\tilde{e}_R}=137.7$ GeV and $m_{\tilde{e}_L}=179.3$ GeV and for different polarizations $P_{e1}=\pm 80\%$, $P_{e2}=\pm 80\%$ and $\lambda_L=\pm 100\%$. The polarizations are denoted as $(sign(P_{e1}),sign(P_{e2}),sign(\lambda_L))$. The shaded area is excluded for $m_{\tilde{\chi}_1^0}>35$ GeV.}
\end{figure}

\begin{figure}[h]
\label{gammagamma}
\centering
\begin{picture}(15,5.2)
\put(0.0,-1.25){\includegraphics{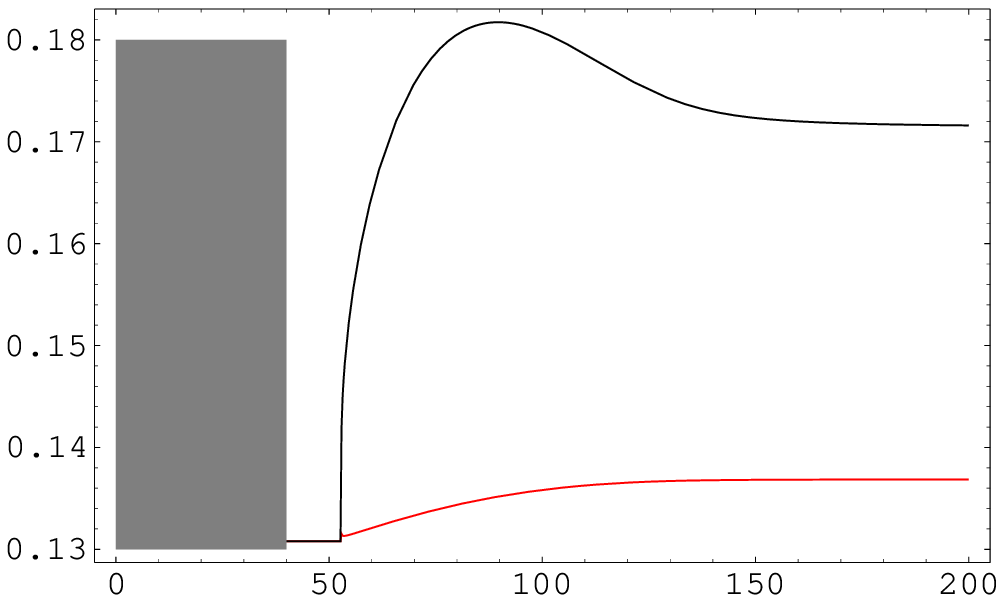}}
\put(6.33,0.05){{\tiny $M_1$/GeV}}
\put(0.05,4.9){{\tiny $\sigma_{ee}$/pb}}
\put(7.9,-1.1){\includegraphics{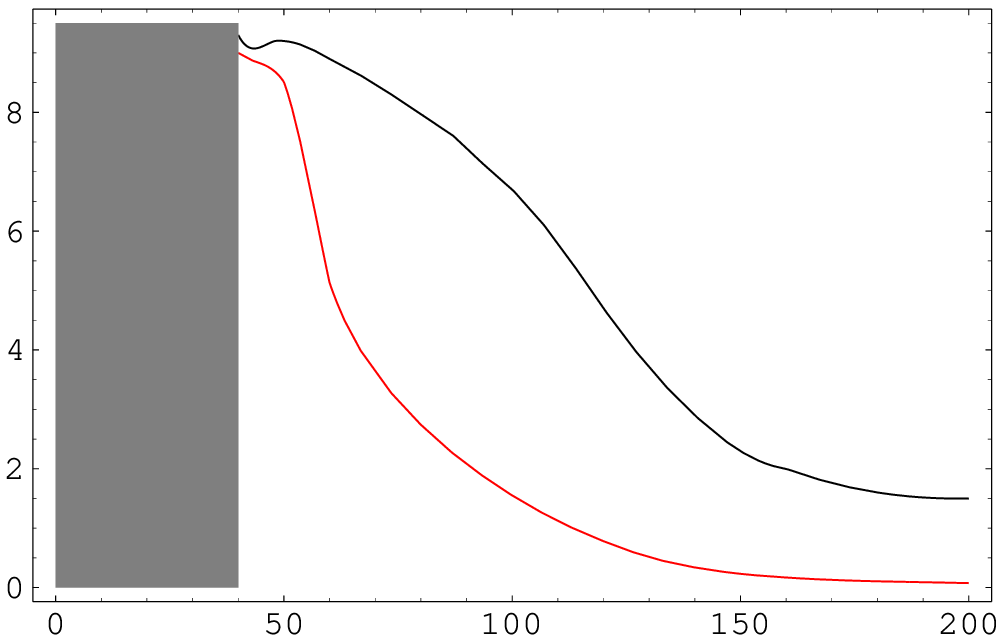}}
\put(13.8,0.05){{\tiny $M_1$/GeV}}
\put(7.8,4.9){{\tiny $-A_{FB}$/\%}}
\put(3.85,4.9){{\tiny (a)}}
\put(11.4,4.9){{\tiny (b)}}
\end{picture}
\caption{Total cross sections $\sigma_{ee}$ (a) and forward-backward asymmetry $A_{FB}$ (b) of $\gamma\gamma \rightarrow \tilde{\chi}_1^+\tilde{\chi}_1^- \rightarrow \tilde{\chi}_1^0\tilde{\chi}_1^0e^+e^-\nu_e\bar{\nu}_e$ for $\sqrt{s_{ee}}=1000$ GeV, $P_{e1}=80\%=-P_{e2}$ and unpolarized laser photons.  Black: $m_{\tilde{\nu}_e}=164.2$ GeV, $m_{\tilde{e}_L}=179.3$ GeV. Red: $m_{\tilde{\nu}_e}=342.5$ GeV, $m_{\tilde{e}_L}=350.0$ GeV.} 
\end{figure}

\section*{Chargino Pair Production in $\gamma\gamma$ Scattering}

In the pair production of charginos in photon collisions only pure QED 
couplings enter. Since $m_{\tilde{\chi}_1^{\pm}}$ is independent of $M_1$,
the process $\gamma\gamma\rightarrow\tilde{\chi}_1^+\tilde{\chi}_1^-\rightarrow\tilde{\chi}_1^0\tilde{\chi}_1^0\nu_e\bar{\nu}_ee^+e^-$ reflects the
$M_1$ dependence of the leptonic three body decay of the charginos 
\cite{tobias}. In fig. 3a the $M_1$
dependence of the convoluted cross section $\sigma_{ee}$ is shown at 
$\sqrt{s_{ee}}=1000$ GeV for unpolarized laser photons and longitudinal
polarizations $P_{e1}=80\%$ and $P_{e2}=-80\%$ of the 
converted electron beams. 
Due to the large cross sections this polarization configuration leads to
small statistical errors for the
forward-backward asymmetry $A_{FB}$ of the decay electrons. For the selectron mass and 
 the sneutrino mass we have choosen $m_{\tilde{e}_L}=179.3$ GeV, 
$m_{\tilde{\nu}_e}=164.2$ GeV  and 
$m_{\tilde{e}_L}=350.0$ GeV, $m_{\tilde{\nu}_e}=342.5$ GeV.
In both cases the $M_1$ dependence of the total cross section (fig. 3a) 
is rather weak. For $M_1<50$ GeV the two body decay 
$\tilde{\chi}_1^{\pm}\rightarrow W^{\pm}\tilde{\chi}_1^0$ dominates by far.
 Also for $M_1>90$ GeV
the cross sections are quite flat, so that only for small selectron and 
sneutrino masses and 
in the  region 50 GeV $<M_1<$ 90 GeV  the cross section shows a moderate
$M_1$ dependence. The
forward-backward asymmetry $A_{FB}$, fig. 3b,
turns out to be more sensitive on $M_1$. If $A_{FB}>2\%$ can be measured with good 
accuracy, it allows to determine $M_1$ up to 150 GeV for the small slepton
masses and up to 90 GeV for the heavy slepton scenario.

\begin{figure}[h]
\label{e-e-}
\centering
\begin{picture}(15,5.3)
\put(0.0,-1.2){\includegraphics{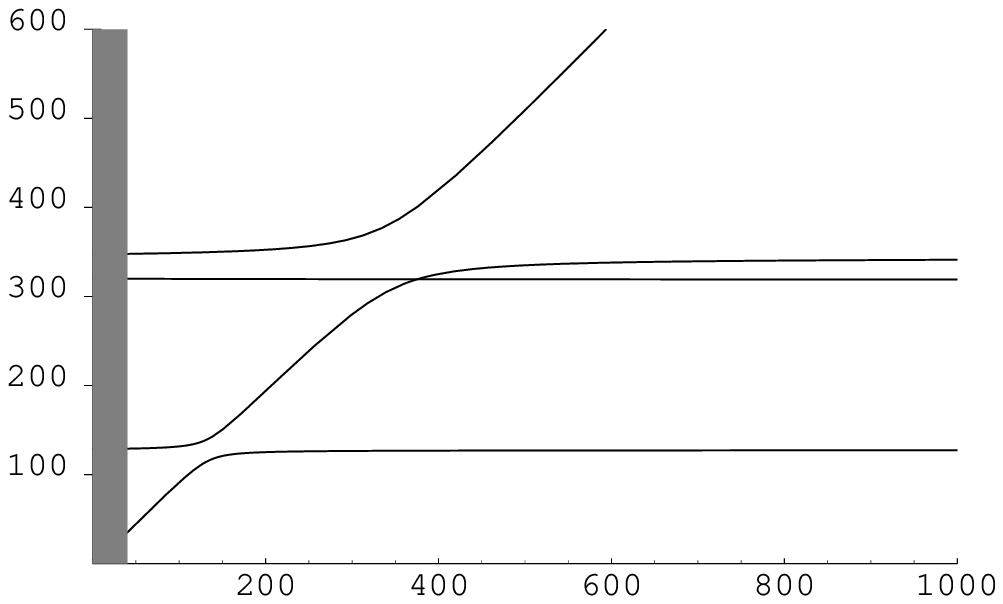}}
\put(6.33,0.05){{\tiny $M_1$/GeV}}
\put(0.05,5.0){{\tiny $\vert m_{\tilde{\chi}_i^0}\vert$/GeV}}
\put(7.6,-1.2){\includegraphics{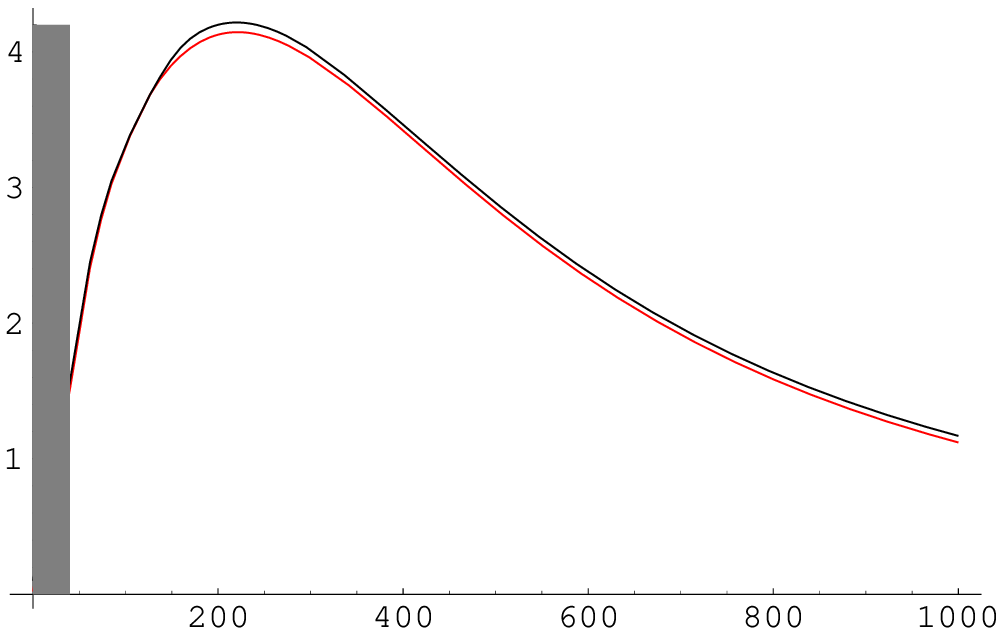}}
\put(13.8,0.05){{\tiny $M_1$/GeV}}
\put(7.6,5.0){{\tiny $\sigma_{ee}$/pb}}
\put(3.85,5.0){{\tiny (a)}}
\put(11.4,5.0){{\tiny (b)}}
\end{picture}
\caption{(a) Neutralino masses as a function of $M_1$; (b) Total cross sections $\sigma_{ee}$ of the process $e^-e^- \rightarrow \tilde{e}_{L/R}^-\tilde{e}_{L/R}^- \rightarrow \tilde{\chi}_1^0\tilde{\chi}_1^0e^-e^-$ for $\sqrt{s}=500$ GeV, longitudinal beam polarizations $P_{e1}=P_{e2}=80\%$ and selectron masses $m_{\tilde{e}_R}=137.7$ GeV, $m_{\tilde{e}_L}=179.3$ GeV. Red: contribution of $\tilde{e}_R\tilde{e}_R$ production and decay.}
\end{figure}

\section*{Selectron Pair Production in $e^-e^-$ Scattering}

In the process $e^-e^- \rightarrow \tilde{e}_{L/R}^-\tilde{e}_{L/R}^- \rightarrow \tilde{\chi}_1^0\tilde{\chi}_1^0e^-e^-$
both selectron production and decay depends on $M_1$. Pair production of 
selectrons proceeds via t- and u-channel exchange 
of all four neutralinos. Since - depending on the respective region - the mass 
and the couplings of at least one of the exchanged neutralinos significantly 
depend on $M_1$ (fig. 4a), one expects that also the cross section of the combined
process is sensitive to $M_1$ over a large region. In fig. 4b the $M_1$ 
dependence of the cross sections for production and subsequent leptonic decay 
of the selectrons is shown at $\sqrt{s}=500$ GeV for longitudinal beam 
polarization $P_{e1}=P_{e2}=+80\%$ and selectron masses 
$m_{\tilde{e}_R}=137.7$ GeV, $m_{\tilde{e}_L}=179.3$ GeV. Also depicted is
the contribution of the production channel $\tilde{e}_R\tilde{e}_R$,
which is by far dominating for right-polarized electrons. 
The cross section shows a significant $M_1$ dependence in the whole region
40 GeV $<M_1<$ 1000 GeV with one ambiguity for $M_1<$ 250 GeV and $M_1>$ 
250 GeV in our scenario. For left-polarized electrons, $P_{e1}=P_{e2}=-80\%$, the
cross sections, dominated by the $\tilde{e}_L\tilde{e}_L$ channel, are about 
one order of magnitude smaller in our scenario. 
They show an additional ambiguity in the
region 40 GeV $<M_1<$ 250 GeV.
Particularly due to the larger cross
sections right-polarized electron beams are more suitable for constraining
$M_1$ in selectron pair production.

\section*{Conclusions}

For a specific mSUGRA inspired scenario we have studied four processes in
different linear collider modes with polarized beams. Total cross sections,
forward-backward asymmetries and polarization asymmetries for production
and leptonic decay of supersymmetric particles are investigated with regard to their dependence on the
gaugino mass parameter $M_1$. Neutralino  production and decay in $e^+e^-$
annihilation is very sensitive to $M_1$ in the studied region. 
Combining measurements of total cross sections,
forward-backward asymmetries and polarization asymmetries with different beam
polarizations resolves ambiguities.
For associated 
LSP-selectron production and decay in $e\gamma$ scattering the cross sections 
are larger and for suitably polarized beams sensitive to $M_1$ up to 300 GeV.
Ambiguities can be resolved by measuring different polarization asymmetries. 
The total cross section for chargino pair production and decay in $\gamma\gamma$ scattering 
is sensitive on $M_1$ only in the small region 50 GeV $<M_1<90$ GeV. 
The forward-backward asymmetry of 
the decay electrons, however,  shows a signifcant $M_1$ dependence up to 150 GeV without
any ambiguities. For selectron pair production and decay in $e^-e^-$ 
scattering the cross section for right polarized beams is 
sensitive to $M_1$ in the large region 40 GeV $<M_1<$ 1000 GeV with one 
ambiguity.
For a more realistic comparison of the four processes,
 Monte Carlo studies including ISR and beamstrahlung 
are necessary.

\end{document}